# Dynamic Competition Between Hubbard and Superexchange Interactions Selectively Localizes Electrons and Holes Through Polarons

Jocelyn L. Mendes[1], Hyun Jun Shin[2], Jae Yeon Seo[2], Nara Lee[2], Young Jai Choi[2], Joel B. Varley[3], and Scott K. Cushing[1*]

[1] Division of Chemistry and Chemical Engineering, California Institute of Technology, Pasadena, California 91125, United States

[2] Department of Physics, Yonsei University, Seoul, 03722, Republic of Korea

[3] Lawrence Livermore National Laboratory, Livermore, California 94550, United States

*Correspondence and requests for materials should be addressed to S.K.C. (email: scushing@caltech.edu)

**ABSTRACT:** Controlling the effects of photoexcited polarons in transition metal oxides can enable the long timescale charge separation necessary for renewable energy applications as well as controlling new quantum phases through dynamically tunable electron-phonon coupling. In previously studied transition metal oxides, polaron formation is facilitated by a photoexcited ligand-to-metal charge transfer (LMCT). When the polaron is formed, oxygen atoms move away from iron centers, which increases carrier localization at the metal center and decreases charge hopping. Studies of yttrium iron garnet and erbium iron oxide have suggested that strong electron and spin correlations can modulate photoexcited polaron formation. To understand the interplay between strong spin and electronic correlations in highly polar materials, we studied gadolinium iron oxide ($GdFeO_3$), which selectively forms photoexcited polarons through an Fe-O-Fe superexchange interaction. Excitation-wavelength-dependent transient extreme ultraviolet (XUV) spectroscopy selectively excites LMCT and metal-to-metal charge transfer transitions (MMCT). The LMCT transition suppresses photoexcited polaron formation due to the balance between superexchange and Hubbard interactions, while MMCT transitions result in photoexcited polaron formation within 250±40 fs. Ab initio theory demonstrates that electron and hole polarons localize on iron centers following MMCT. In addition to understanding how strong electronic and spin correlations can control strong electron-phonon coupling, these experiments separately measure electron and hole polaron interactions on neighboring metal centers for the first time, providing insight into a large range of charge-transfer and Mott-Hubbard insulators.

## INTRODUCTION

Polaron formation in transition metal oxide materials has been widely explored for its effects on ground state transport in correlated materials and solar energy processes like photocatalysis.[1–5] In highly polar materials, the coupling of photoexcited charge carriers to phonons in the lattice localizes carriers in the form of polarons.[6] Parameters to dynamically control laser-driven strong electron-phonon coupling is an ongoing search. Understanding and controlling the underlying physics that determines the formation of polarons in materials could provide meaningful improvements to a wide range of material classes, from organic electronics to solar energy systems and exotic quantum phases.[7,8]

Hematite ($\alpha$-$Fe_2O_3$) and other iron oxides have been extensively explored for their photoexcited polaron formation dynamics.[9–12] Most of these materials are charge-transfer insulators, where the valence band maximum (VBM) is dominated by O 2p orbitals, and the conduction band minimum (CBM) is dominated by metal 3d orbitals. An above-band gap photoexcitation induces a ligand-to-metal charge transfer (LMCT) transition from oxygen to the metal center. The electronic structures of metal oxides also span intermediate and Mott-Hubbard insulators, where the VBM is dominated by either mixed oxygen/metal orbitals or entirely metal orbitals, respectively. These differing electronic structures pose a unique route for modulation of carrier localization by polaron formation following photoexcitation.

Almost uniformly across iron oxides, photoexcited small polaron formation occurs within a few hundred femtoseconds of the charge transfer transition.[10,13,14] A subsequent expansion of Fe-O bond lengths and localization of electron density on polaronic iron sites has been measured using a variety of X-ray spectroscopies. Recent work exploring the tunability of polaronic properties has been demonstrated using structural distortions and strong electron and spin correlations. For example, rare-earth orthoferrite ErFeO$_3$ was found to have weaker polaronic binding energies because the polaron formation rate was slowed by strong electronic correlations.[12] Additional work comparing hematite and Y$_3$Fe$_5$O$_{12}$ (YIG) finds that strong spin selective dynamics reduce polaron formation in YIG and increase its photocatalytic efficiency by an order of magnitude with respect to hematite.[15] Transient XUV and X-ray spectroscopies, however, have yet to be applied to the simultaneous measurement of electron and hole polarons in intermediate and Mott-Hubbard insulating iron oxides.

Here, we employ excitation-wavelength-dependent transient extreme ultraviolet (XUV) reflection spectroscopy to explore the relation between superexchange and Hubbard interactions in influencing strong carrier localization through polarons in single-crystal rare-earth orthoferrite GdFeO$_3$. This intermediate insulator (mixed O 2p/Fe 3d VBM) is a model material system for many perovskite Mott-Hubbard insulators due to its crystal structure.[16,17] GdFeO$_3$ and other rare-earth orthoferrites have been explored as candidates for photoelectrochemical oxygen reduction and evolution, as well as magneto-optical and spintronic applications due to their demonstrated multiferroicity.[18,19] Through transient XUV spectroscopy, we find that LMCT from O 2p to Fe 3d orbitals results in the suppression of photoexcited polaron formation, while a metal-to-metal charge transfer (MMCT) induces photoexcited polaron formation. The MMCT enables polarons by superexchange across an Fe-O-Fe bond, resulting in hole and electron polarons localized on neighboring iron centers, and a reduced radiative lifetime (<100 ps for MMCT versus 1.2±0.4 ns for the LMCT). Polaron formation is suppressed following a higher energy LMCT due to the balance of on-site Coulomb repulsion (U) from oxygen dominated valence orbitals and superexchange in the final spin state. While thermalization of these carriers occurs in ~350 fs, the final spin state from the LMCT prevents the polaron formation. Ab initio density functional theory (DFT) and the Bethe-Salpeter equation (BSE) simulate the photoexcited XUV dynamics. Polaron-induced lattice distortions and charge localization within a defect supercell approach were assessed with different levels of electronic structure theory, ranging from DFT incorporating on-site Hubbard U parameters with variational polaron self-interaction-corrected (pSIC) total-energy functional models to hybrid functionals with different fractions of exact-exchange. The results provide new insight into tuning polaronic properties via strong electronic and spin correlations for transition metal oxide materials, particularly identifying superexchange interactions as a new dynamical carrier localization and polaron design parameter.

## EXPERIMENTAL
### Synthesis of single-crystalline GdFeO$_3$
Single crystals of GdFeO$_3$ were synthesized using the flux method, employing PbO, PbF$_2$, PbO$_2$, and B$_2$O$_3$ as fluxes in a high-temperature furnace. A stoichiometric mixture of Gd$_2$O$_3$ and Fe$_2$O$_3$ powders was thoroughly combined with the flux compounds and placed in a platinum crucible. The mixture was heated to 1250 °C for 16 h to ensure complete dissolution. Then, it was cooled slowly to 850 °C at a rate of 2 °C/h, and further cooled to room temperature at a rate of 100 °C/h. This process yielded large cuboid shaped GdFeO$_3$ crystals, with individual crystals obtaining lengths of up to 1 cm on a side. Characterization of the GdFeO$_3$ single crystals can be found in the Supporting Information.

### Transient extreme ultraviolet (XUV) spectroscopy
For both photoexcitation wavelengths, the pump pulses are ~50-fs with a fluence of ~11 mJ/cm$^2$ and are time-delayed with respect to the probe pulse. In the case of 800 nm photoexcitation, this results in a photoexcited carrier density of ~7.3x10$^{20}$ cm$^{-3}$, and for 400 nm excitation a carrier density of ~3.7x10$^{20}$ cm$^{-3}$, details of the carrier density calculation can be found in the Supporting Information. The DFT calculated unit cell of GdFeO$_3$ has a volume of ~2.4x10$^{-22}$ cm$^3$, and given that each unit cell contains 4 iron atoms, there are approximately ~1.6x10$^{22}$ Fe atoms/cm$^3$. This would result in ~4.5% and ~2.3% of iron atoms excited following 800 nm and 400 nm photoexcitation, respectively. The XUV probe pulse is generated by high harmonic generation in Ar gas from a few-cycle white light pulse. The transient extreme ultraviolet reflection experiment employs a 10° grazing incidence geometry, which results in a ~2 nm penetration depth.[20] The Fe M$_{2,3}$ edge at ~54 eV corresponds to transitions from the 3p$_{3/2,5/2}$ core levels into 3d core levels. The change in transient reflection following photoexcitation is defined by ΔOD = -log$_{10}$(I$_{pump\ on}$/I$_{pump\ off}$). Additional details about the transient extreme ultraviolet spectrometer can be found in the Supporting Information.

**Ab initio polaron and core-level spectra modeling**

The theoretical framework we employ to fully analyze the transient core-level spectra has been described in detail previously.[21] We compute both the ground state X-ray absorption at the Fe $M_{2,3}$ edge in $GdFeO_3$ and differential effects to the core-level $GdFeO_3$ spectra following excitation-wavelength-dependent photoexcitation and polaron formation. These effects are calculated with an ab initio theoretical approach which employs density functional theory and the Bethe-Salpeter equation (DFT+BSE). The DFT+BSE approach employs DFT using the Quantum ESPRESSO package[22,23] and the existing OCEAN code[24–26] (Obtaining Core-level Excitations using Ab-initio calculations and the NIST BSE solver). We have modified the BSE to use excited state distributions to determine transient changes in the XUV spectra. We first calculate the band structure of $GdFeO_3$ using DFT. Then, the BSE is solved to obtain the core-valence exciton wavefunctions and transient XUV spectra can be plotted from the complex dielectric function calculated by OCEAN. A modification to the OCEAN code, discussed previously,[21,27] enables the calculation of excited state dynamics and is employed to model the initial 800 nm MMCT and 400 nm LMCT states.

Polaronic distortions are modeled using a DFT+BSE approach where we apply a semi-empirical distortion to an $FeO_6$ octahedra and calculate its resulting Fe $M_{2,3}$ edge spectra. Polaronic distortions calculated using this technique are termed DFT+BSE. The polaronic distortions are modeled independently of the different charge transfer transitions, but it is implied that only MMCT could induce hole polarons on iron centers. We further model polaronic lattice distortions using a defect supercell approach with hybrid and polaron self-interaction corrected exchange-correlation functionals (HSE06 and pSIC) as implemented in the VASP code.[28–32] These functionals either incorporate a fraction of Hartree-Fock (HF) exact exchange to improve self-interaction errors and a description of charge localization (e.g. hybrid functionals) that depends on an empirical parameter ($\alpha$, the HF mixing parameter), or employ a variational approach from a closed-shell system for a parameter-free ab initio methodology for calculating polaron properties that removes the sensitivity to empirical parameters like $\alpha$ or Hubbard U parameters.[29] These methods were used to identify structural distortions and relative energetics associated with electron and hole polaron formation at different sites in $GdFeO_3$ and to assess their sensitivity to empirical parameters in the calculations. We generally refer to the polaron-containing structures and subsequent X-ray spectra obtained via these approaches (hybrid functionals and/or with the pSIC approach) as pSIC+DFT+BSE. Additional details about ab initio modeling of photoexcited XUV dynamics can be found in the Supporting Information section.

## RESULTS AND DISCUSSION

**Defect Supercell Calculations of Polarons**

$GdFeO_3$ has an orthorhombic symmetry in the *Pbnm* space group, as shown in Figure 1A.[17] The $FeO_6$ octahedra are strained due to the large rare-earth atom in the crystal structure, resulting in a 154° Fe-O-Fe bond angle.[33] Strong electron correlations introduced by both the Fe d and Gd f orbitals enable a mixed Fe d/O p band at the VBM and a metal d band at the CBM as shown in Figure 1B. This contrasts with previously measured iron oxides whose density of states follows a charge-transfer insulating structure where the VBM is dominated by O 2p orbitals.[10,21] DFT+U is used to accurately describe the extent of carrier localization and on-site repulsion in $GdFeO_3$.[17] Magnetization studies of $GdFeO_3$ have demonstrated that it is a canted, G-type antiferromagnet (AFM) below its Néel temperature ($T_N$) of ~650 K, resulting in a weak ferromagnetism due to the Dzyaloshinskii-Moriya (DM) interaction.[34–36] The $Fe^{3+}$ ions in the $GdFeO_3$ take a high-spin (HS) electronic configuration in their ground state, and the octahedral tilts of adjacent $FeO_6$ octahedra are measured to have strong antisymmetric exchange interactions (superexchange).[37,38]

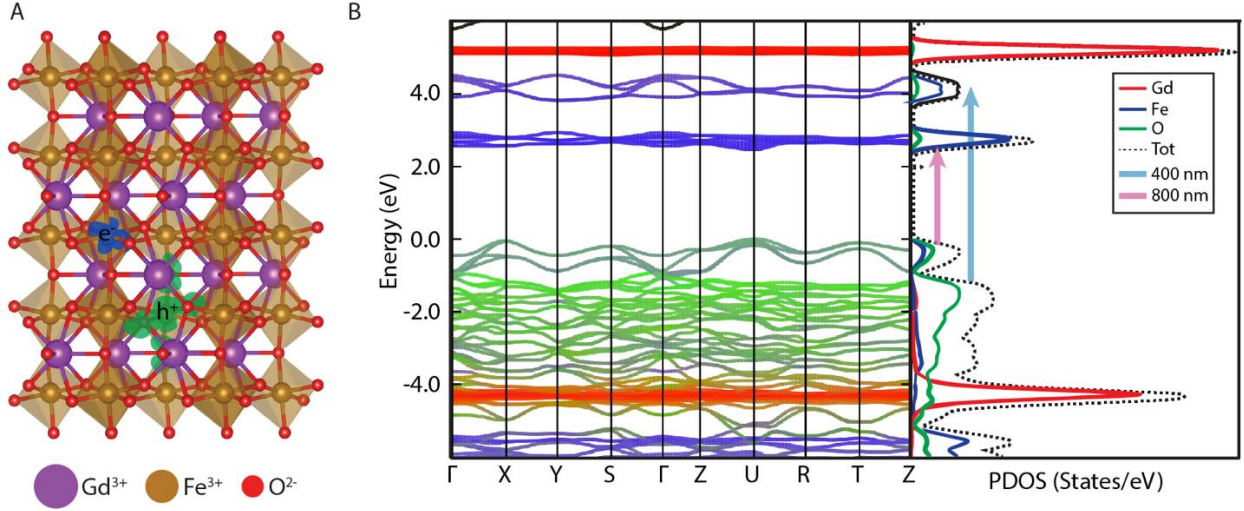

**Figure 1.** (A) GdFeO$_3$ crystal structure with electron (blue) and hole (green) polaron charge densities as calculated with exchange correlation functionals that correct for self-interaction errors, referred to as defect supercell calculations. The electron and hole polarons are both predicted to be centered on iron atoms. (B) Calculated electronic band structure and projected density of states (PDOS) with the HSE06 hybrid functional with 15% exact exchange for the studied AFM ground state, including tentative assignments of band transitions associated with the 800 nm (pink arrow) and 400 nm (blue arrow) excitations that lead to the polaron formation. The bandgap is overestimated due to deficiencies associated with DFT of transition metals, but the band structure and contributions of relative orbitals is preserved. The excitation arrows are adjusted accordingly to showcase the transition densities expected based upon the experimentally measured band gap.

Figure 1B shows the two photoexcitation pathways used in the excitation-wavelength-dependent XUV measurements. In the case of the 400 nm (3.1 eV) photoexcitation, carriers are excited from lower lying valence bands that are dominated by O 2p orbitals into higher energy Fe 3d and Gd 6f orbital dominated conduction bands. Photoexcitation with 400 nm light follows a LMCT mechanism. The lower energy 800 nm (1.55 eV) photoexcitation induces transitions from the VBM, which are mixed in O 2p and Fe 3d orbital character, into lower energy conduction bands dominated by Fe 3d orbitals. The mixed nature of the VBM in GdFeO$_3$ poses an opportunity for both MMCT and LMCT to occur.[39] These excitation energies provide different possible pathways for polaron formation.

We employ a defect supercell approach with hybrid and polaron self-interaction corrected exchange-correlation functionals (HSE06 and pSIC) to calculate ab initio polaronic distortions in GdFeO$_3$. Further details of the calculations can be found in the Experimental section and the Supporting Information. These methods identify structural distortions and relative energetics associated with electron and hole polaron formation at different sites in GdFeO$_3$ and assess their sensitivity to empirical parameters in the calculations. The charge localization of both electron and hole polarons from defect supercell calculations (Fig. 1A) predicts that the electron polaron localizes on one single iron site. The hole polaron is less localized than the electron polaron, however, its charge density is predicted to be centered over an iron site as well. A polaron formed following MMCT would induce both electron and hole polaron formation on iron centers. We hypothesize that the lack of an oxygen-centered polaron from the mixed orbital VB means that the LMCT may not lead to polaron formation.

## Hubbard-Holstein Model of Polaron Formation

To understand the defect supercell predicted localization of electron and hole polarons on iron centers in GdFeO$_3$, we must understand the conditions that make polaron formation favorable in a material system. The model Hamiltonian that follows the two-site Hubbard-Holstein model in Equation 1 presents the relevant properties that dictate polaron formation:

$$H = -t_{ij} \sum_{\langle ij \rangle \sigma} (c^\dagger_{i\sigma} c_{j\sigma} + H.C.) + U \sum_i n_{i\uparrow} n_{i\downarrow} + J \sum_{ij} \vec{S_i} \cdot \vec{S_j} + \omega_0 \sum_i b^\dagger_i b_i + g \sum_{i,\sigma} n_{i\sigma}(b_i + b^\dagger_i) \quad (1)$$

Where $t_{ij}$ is the hopping integral, $c^\dagger_{i\sigma}$ and $c_{j\sigma}$ are the creation and annihilation operators, respectively, with spin, $\sigma$, for an electron at site $i$ and $j$.[40–45] $U$ is the on-site Coulombic repulsion, $n_i$ is the number operator, $g$ corresponds to the electron-phonon coupling strength, and $\omega_0$ refers to the phonon frequency. $b^\dagger_i$ and $b_i$ are the creation and annihilation operators, respectively, for phonons at site $i$. $J$ denotes the superexchange interaction between neighboring metal sites.[46] The Hubbard-Holstein Hamiltonian is commonly downfolded to a t-J model.[47,48] Here we consider the Hubbard-Holstein model for simplicity, but a t-J Hamiltonian could be applied under certain coupling regimes.

Polarons exist in a strongly electron-phonon coupled regime, as carriers are localized by strong interactions with the lattice. Alternatively, when electron-phonon coupling is sufficiently weak, carriers may conduct freely in materials and polaron formation is suppressed. This indicates that for polaron formation to be favorable $t \ll g$, while polaron formation would be weak or suppressed when $t \gg g$. In a $U \gg t$ regime at half filling, $J \approx 4t^2/U$, and for an antiferromagnet $t_{eff} \propto J$ and $t_{eff} \propto 1/U$.[46,49,50] We note that this is an oversimplification, as U and J are changing many body wavefunctions. DMFT work to study the interplay of electron-phonon coupling and the superexchange interaction find that as $g$ increases, $J$ decreases, meaning that superexchange is suppressed in a strongly electron-phonon coupled regime.[41] On-site Coulombic repulsions are inversely related to $J$, and thus, $J$ is also expected to be suppressed in a regime where $U$ is sufficiently large.[46]

The defect supercell calculations of $GdFeO_3$ predict that both electron and hole polarons will form on metal sites (Fig. 1A). This contradicts previous measurements of polaron formation, in which the hole polaron would be localized on a ligand site and the electron polaron would be localized on a metal site. We hypothesize that this is primarily due to the different accessible excitation manifolds within this intermediate insulator. $GdFeO_3$ is expected to have strong $J$ in its ground state due to the AFM configuration of adjacent Fe sites and the tilted $FeO_6$ octahedra.[38] However, the strong $U$ associated with an MMCT will decrease charge hopping and increase the likelihood of polaron formation. On the other hand, LMCT would originate from a ligand with weak $U$, which would be associated with weakened electron-phonon coupling. In this case, charge hopping would be expected to be strong, and polaron formation will be suppressed.

**Suppressed Polaron Formation Following LMCT**

XUV spectroscopy has been employed extensively for the measurement of photoexcited polaron formation dynamics because it can evaluate ultrafast, element-specific electronic-structural dynamics.[9,10,14] Spectrally, polaron formation in iron oxide materials has been characterized in the XUV as a shift to higher energies at the Fe $M_{2,3}$ edge. As carriers localize on the iron center, more energy is required to induce the core-to-valence 3p to 3d transition at the iron site. This has held for charge-transfer polarons across Fe-O materials with dynamics that agree well with transient X-ray diffraction and ultrafast optical techniques.[51,52]

Transient XUV reflectivity at the Fe $M_{2,3}$ edge of $GdFeO_3$ pumped with 400 nm (3.1 eV) light is shown in Figure 2A. The Fe $M_{2,3}$ edge is characterized immediately after photoexcitation by a negative absorption feature (blue) centered around 54 eV. This spectral feature can be described by a reduction in the population of the $Fe^{3+}$ ground state (Fig. S4) as electrons are excited into Fe d orbitals in the $GdFeO_3$ conduction bands.[10] Transient increases in absorption are present in the 400 nm pumped spectra (Fig. 2A) from ~52.6 – 52.9 and ~55.6 – 58.0 eV. Despite no spectral shift to higher energies being observed at the Fe $M_{2,3}$ edge following 400 nm photoexcitation, there is a decrease in intensity over time (Fig. 2A). Figure 2C provides an exponential fit of this decrease in intensity, resulting in a rate constant of $\tau_{400\,nm\,hot\,carriers}$ = 360±70 fs, which we attribute to the cooling rate of the hot carriers, in line with XUV studies of photoexcited carrier cooling.[20,53–55]

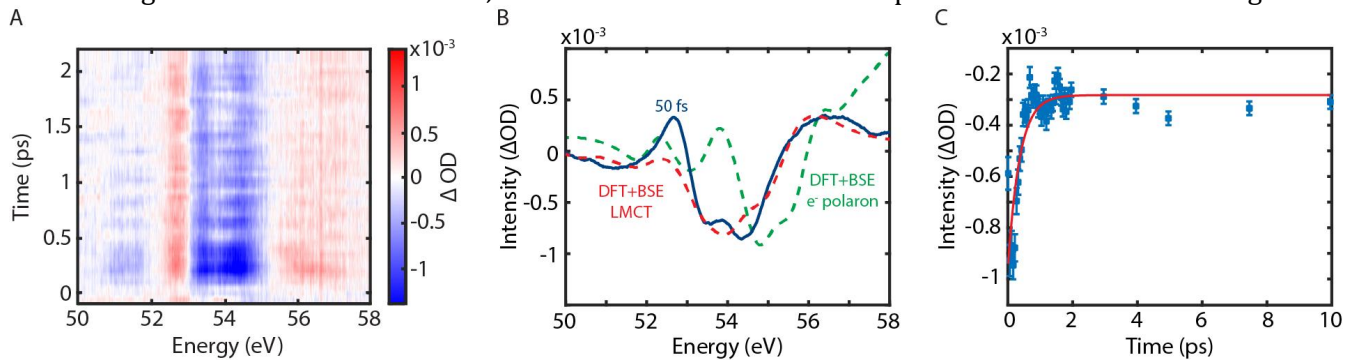

**Figure 2.** (A) Transient XUV spectra following 400 nm photoexcitation. (B) Experimental lineout from 50 fs after 400 nm photoexcitation (blue) compared to DFT+BSE theory lineouts that model the 400 nm LMCT state (red) and an electron polaron state on an iron atom (green) show agreement with a LMCT state following 400 nm photoexcitation. (C) Fitting of the thermalization at the Fe $M_{2,3}$ edge following 400 nm photoexcitation, averaged over the Fe $M_{2,3}$ edge feature.

Due to the strong contribution of angular momentum to the core-level transition Hamiltonian at the Fe $M_{2,3}$ edge, increases (red) and decreases (blue) in absorption in the XUV spectrum do not directly relate to electron and hole energies following photoexcitation, but instead relate to changes in oxidation state, phonon modes, and other structural distortions such as small polarons.[10,12] To understand the origin of the spectral features at the Fe $M_{2,3}$ edge of $GdFeO_3$, we use a density functional theory and Bethe-Salpeter equation (DFT+BSE) approach to model the excited state X-ray edge dynamics. The method uses the Quantum ESPRESSO[22,23] and the OCEAN[24–26] (Obtaining Core Excitations from the Ab initio electronic structure and the NIST BSE solver) codes. Our defect supercell approach with hybrid and polaron self-interaction corrected exchange-correlation functionals (HSE06 and pSIC) calculates ab initio polaronic distortions in $GdFeO_3$ which are applied to our DFT+BSE framework (pSIC+DFT+BSE). We also model charge transfer and semi-empirical polaronic states using this DFT+BSE approach. The theoretical methods applied here are discussed in more detail in the Experimental section and Supporting Information.

Figure 2B compares an experimental lineout following 400 nm photoexcitation with DFT+BSE theory modeled 400 nm LMCT and electron polaron states. There is good agreement between experimental lineouts at varying time delays following the 400 nm pump and the modeled 400 nm LMCT state. There is poor agreement between the 400 nm experimental lineouts and the DFT+BSE modeled polaron state that shifts to higher energy. Further, the increases in absorption following 400 nm photoexcitation (Fig. 2A) from ~52.6 – 52.9 and ~55.6 – 58.0 eV are described by high-spin $Fe^{2+}$ states forming as LMCT from the oxygen to iron atoms occurs (Fig. 3A) using ligand-field multiplet theory.[56] A low-spin (LS) $Fe^{2+}$ state does not agree with the experimentally measured spectrum (Fig. 3A).

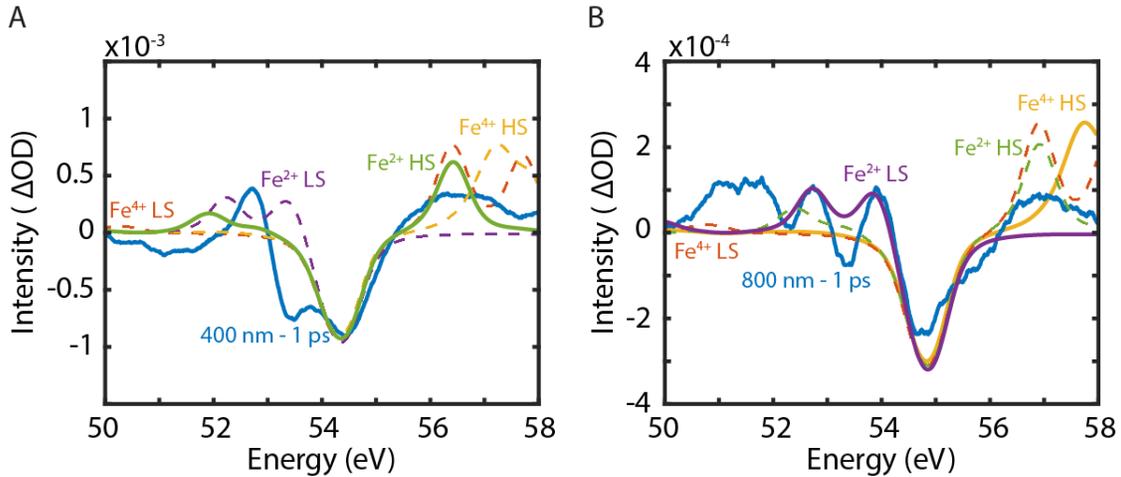

**Figure 3.** Experimental (blue) 400 nm LMCT state (A), 800 nm polaron and trapped $Fe^{2+}$ low-spin state (B). CTM4XAS calculated $Fe^{4+}$ high-spin state (yellow), $Fe^{4+}$ low-spin state (red, dashed), $Fe^{2+}$ high-spin state (green), $Fe^{2+}$ low-spin state (purple). Each oxidation and spin state are plotted for comparison, but the solid lines refer to the assigned CTM4XAS calculated spin states for both experimental spectra.

The photoexcited LMCT state occurs on the same timescale as the pulse width and is present up to our temporal detection limit of 1 ns (Fig. S5B). The non-radiative recombination is, therefore, longer than our instrument's 1 ns temporal limit. An exponential fit of the radiative lifetime of $GdFeO_3$ gives a time constant of $\tau_{400\ nm\ radiative}$ = 1.2±0.4 ns (Fig. S3), but we cannot confirm if the radiative lifetime is from defects or polarons based on our measurement limits. Further details of fluorescence lifetime measurements can be found in the Supporting Information.

If we consider the hot carrier thermalization from Figure 2C, with a time constant of $\tau_{400\ nm\ hot\ carriers}$ = 360±70fs and compare it to the DFT calculated phonon band dispersion for $GdFeO_3$ we can approximate the energetic extent of carrier cooling to the CBM. $GdFeO_3$ has its highest frequency phonon mode occurring around ~17 THz, which corresponds to ~70.3 meV.[57] At a first-order approximation, we can apply Equation 2 to calculate the amount of energy the hot carriers cool by:

$$\Delta E_{hot\ carriers} = \frac{\tau_{400\ nm\ hot\ carriers}}{\tau_{e-ph}/2} * E_{ph} \quad (2)$$

where $E_{ph}$ is the highest frequency phonon mode and $\tau_{e\text{-}ph}$ is the electron-phonon scattering rate, which can also be approximated by the highest frequency phonon mode in the phonon band dispersion. The factor of two corresponds to the half phonon period frequency on which electron-phonon scattering occurs, a common approximation in literature.[58,59] This results in ~750 meV of thermalization. The experimentally measured direct band gap of single crystal $GdFeO_3$ is ~1.4 eV (Fig. S1), so electrons excited by the 400 nm pump would be ~2.2 eV above the VBM. This would indicate that excitations occur from ~1 eV below the VBM and the PDOS demonstrates that oxygen orbitals dominate those valence bands (Fig. 1B), which supports that an LMCT occurs following 400 nm photoexcitation.

Our DFT+BSE, ligand-field multiplet theory, and fluorescence lifetime measurements, therefore, support that LMCT following 400 nm photoexcitation forms a high-spin $Fe^{2+}$ excited state, but strong coupling to phonons that would induce a polaronic lattice distortion is not present. The lack of a spectral shift to higher energies following 400 nm photoexcitation suggests that polaron formation is suppressed or so weak that it does not influence the transient spectra, and does not compete with the free carriers and their thermalization during the LMCT step (Fig. 2B).[10,12]

**Polaron Formation Following MMCT**

Transient XUV reflectivity at the Fe $M_{2,3}$ edge of $GdFeO_3$ following 800 nm (1.55 eV) photoexcitation is shown in Figure 4. Like 400 nm pumped transient XUV reflectivity spectra, the Fe $M_{2,3}$ edge following an 800 nm pump is characterized immediately after photoexcitation by a negative absorption feature (blue) centered around 54 eV. Unlike in the 400 nm spectra, the negative absorption feature pumped with 800 nm excitation has a spectral blueshift at the Fe $M_{2,3}$ edge from ~52.8 – 55.8 eV (blue) that begins shortly after photoexcitation. This spectral shift is most prominent from ~54.4 – 56.0 eV and has a time constant of 250±40 fs (Fig. S6) with an energy shift of 420±150 meV. Following this spectral shift, a peak splitting results in a positive absorption feature (red, ~53.5 – 54.3 eV) within the Fe $M_{2,3}$ edge in the 800 nm pumped spectra. Additionally, following 800 nm photoexcitation, an increase in absorption (Fig. 4) that appears from 50.3 – 52.8 eV and 56.3 – 58.0 eV is measured.

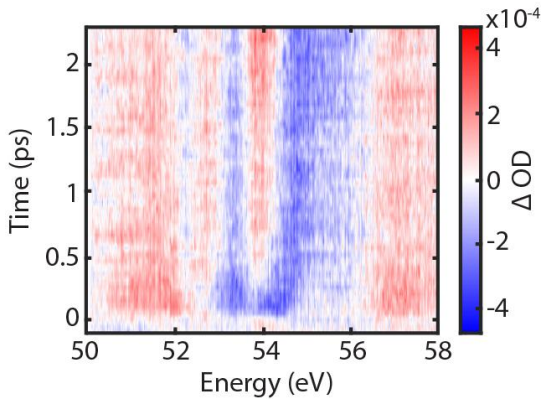

**Figure 4.** Transient XUV reflection-absorption spectrum following 800 nm photoexcitation.

The spectral lineout 50 fs after 800 nm photoexcitation in Figure 5A (blue) agrees well with DFT+BSE theory (red) for a modeled 800 nm MMCT state. The DFT calculated PDOS (Fig. 1B) demonstrates that this excitation originates from mixed O 2p/Fe 3d valence bands and, therefore, could be either LMCT or MMCT in character. We find that the positive absorption features (Fig. 4) at lower energies (50.0 – 52.8 eV) and higher energies (56.3 – 58.0 eV) in the 800 nm pumped spectra are associated with the creation of low-spin $Fe^{2+}$ and high-spin $Fe^{4+}$ states, respectively (Fig. 3B). The formation of high-spin $Fe^{4+}$ states and low-spin $Fe^{2+}$ states suggests that MMCT transitions are occurring between iron atoms following photoexcitation.[39] Section S5, in the Supporting Information, further discusses ligand-field multiplet theory to model these spin states.

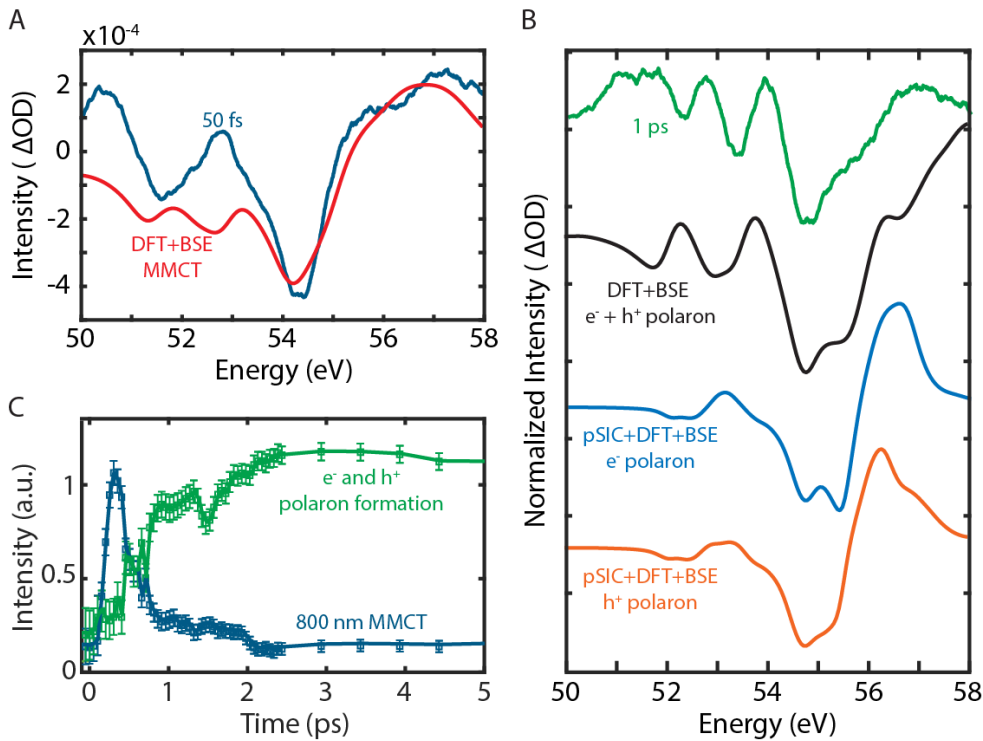

**Figure 5.** (A) Experimental lineout (blue) 50 fs after 800 nm photoexcitation and DFT+BSE theory calculated 800 nm MMCT state (red). (B) The polaron state that forms within a few hundred femtoseconds of 800 nm photoexcitation is modeled using DFT+BSE theory (black) for an electron and hole polaron and using the defect supercell pSIC+DFT+BSE theory for electron (blue) and hole (orange) polarons and is shown experimentally (green, 1 ps after 800 nm photoexcitation). Singular value decomposition (C) of the experimental 800 nm pumped transient spectra reveals the kinetics of photoexcited electron and hole polaron formation in $GdFeO_3$.

Figure 5B shows agreement between experiment after the blueshift (green), DFT+BSE theory for a convolved electron and hole polaron (black), and pSIC+DFT+BSE theory for both electron (blue) and hole (orange) polarons at neighboring iron sites. The pSIC+DFT+BSE calculated electron and hole polarons occur at similar energies, suggesting that the electron and hole polarons are convolved spectrally. This confirms agreement between semi-empirical DFT+BSE and ab initio pSIC+DFT+BSE theory, suggesting that different levels of theory can similarly describe the spectral shifts and features associated with convolved electron and hole polaron formation at the Fe $M_{2,3}$ edge. Singular value decomposition (SVD) in Figure 5C illustrates the transition of the MMCT state to the convolved electron and hole polaron state in the spectra. Figure S7 in the Supporting Information provides further details of the SVD analysis. The SVD plot demonstrates that the MMCT state transitions to the convolved electron and hole polaron state within the first picosecond following 800 nm photoexcitation. Long timescale 800 nm XUV spectra (Fig. S5A) reveal that the spectral shift lasts out to our instrument's temporal detection limit, and a fit of the exponential decay reveals a time constant of 150±30 ps.

**Polaron Formation and Fe-O-Fe Superexchange**

In addition to forming photoexcited electron and hole polarons on iron centers following 800 nm photoexcitation, a strong spectral splitting of the Fe $M_{2,3}$ edge feature occurs following polaron formation. The spectral splitting in the 800 nm spectra (Fig. 4) results in an increase in absorption at ~53.5 – 54.3 eV that arises 330±50 fs following photoexcitation. In the transient XUV spectra following 400 nm photoexcitation (Fig. 2A), there is spectral splitting that occurs at the Fe $M_{2,3}$ edge (53.8 eV), but it is weaker in intensity than the 800 nm photoexcitation, does not result in a positive signal, and does not change in intensity over time.

The spectral splitting can be attributed to $t_{2g}$ and $e_g$ crystal field splitting expected from the $GdFeO_3$ $Fe^{3+}$ high-spin ground state (Fig. S4).[37,60] This high-spin electronic configuration is preserved following 400 nm LMCT (Fig. 3A), as demonstrated by the lack of change in spectral splitting between the ground state spectra and excited state 400 nm pumped spectra (Fig. 2A and Fig. S4). Additionally, the increases in absorption following 400 nm photoexcitation (Fig. 2A) result in the formation of an $Fe^{2+}$ high-spin state, that appears as increases in absorption at ~52.6 – 52.9 and ~55.6 – 58 eV (Fig. 3A). This indicates that the suppression of polaron formation following 400 nm photoexcitation is related to the preservation of the spin state of the iron atoms following LMCT. Further, the LMCT state does not induce polaron formation because the electron density added to the iron atom originates from an oxygen atom. The ground state DFT predicted bond angle between Fe-O-Fe octahedra (~147.1°) is highly conducive to superexchange. The Hubbard interaction is also weak on the oxygen atom. In this regime, the effective charge hopping integral, $t_{eff}$, remains large, suppressing polaron formation in the final, thermalized high spin state.

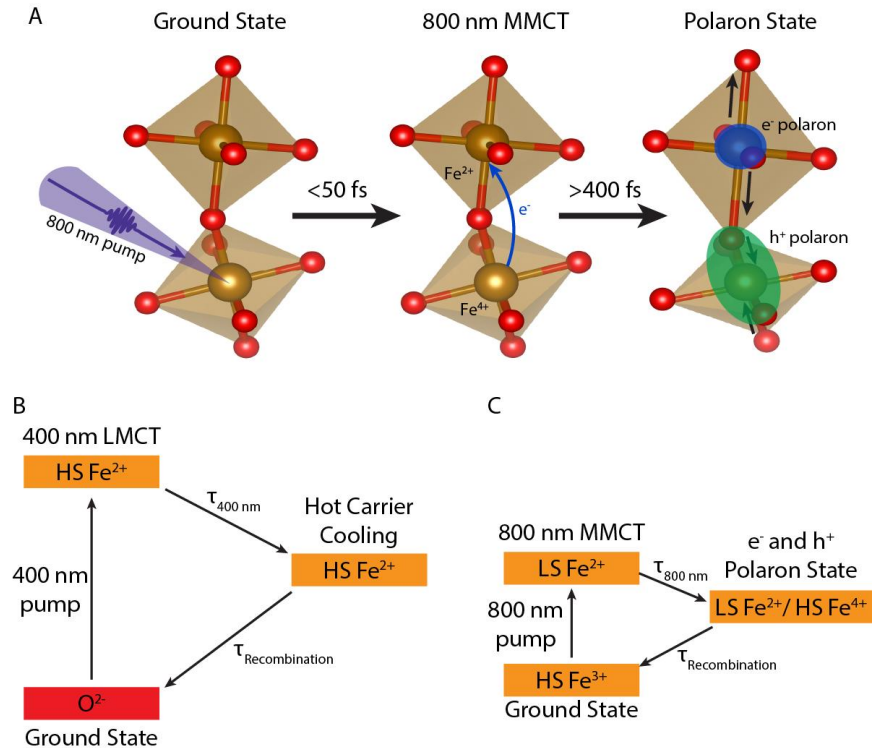

**Figure 6.** (A) Schematic of superexchange across the Fe-O-Fe bond that gives rise to spin crossover on $Fe^{2+}$ centers during photoexcited electron and hole polaron formation. Within the first 50 fs, the 800 nm MMCT state results in a charge transfer from one iron center to another. After >400 fs, electron and hole polaron formation results in an expansion and contraction along the Fe-O-Fe bond length, that modulates its bond angle, inducing Jahn-Teller distortions and trapping the high-spin $Fe^{2+}$ state. Diagram of the excitation pathways and spin manifolds present following both 400 nm (B) and 800 nm excitation (C). The fitting for the time constants for both hot carrier cooling following 400 nm LMCT ($\tau_{400\,nm}$ = 360±70 fs) and polaron formation and superexchange following 800 nm MMCT ($\tau_{800\,nm}$ = 250±40 fs) are discussed further in the SI.

In the case of 800 nm photoexcitation, the strong electron-phonon coupling induced by photoexcited electron and hole polarons influences the dynamics of spectral splitting at the Fe $M_{2,3}$ edge. This is evidenced by the $t_{2g}$ and $e_g$ crystal field splitting in the 800 nm pumped spectra (Fig. 4) experiencing a transient increase in absorption at ~53.5 – 54.3 eV.[61] As demonstrated in the DFT+BSE and pSIC+DFT+BSE modeled spectra (Fig. 5B), the $Fe^{2+}$ and $Fe^{4+}$ centers created following MMCT begin forming electron and hole polarons, respectively. The hole polaron charge density is distributed across the oxygen and $Fe^{4+}$ iron atom, while the electron polaron charge density is localized on the $Fe^{2+}$ site (Fig. 1A). This is supported by ligand-field multiplet theory modeling of both high-spin and low-spin configurations of $Fe^{2+}$ and $Fe^{4+}$ (Fig. 3B), that agrees well with a low-spin $Fe^{2+}$ configuration following photoexcitation. The $Fe^{4+}$ feature at higher energies does not experience this spectral splitting, which suggests that the iron that forms the hole polaron does not experience this spin crossover and remains in a high-spin configuration. This agrees with reports of delocalization, or "sharing", of holes across a metal and ligand participating in superexchange.[62,63]

The distortion induced by the polarons is anisotropic (Fig. S10) and results in a Jahn-Teller distortion of the axial Fe-O ligands between the Fe-O-Fe bond length that is participating in polaron formation (Fig. 6A). The Jahn-Teller effect has been demonstrated to induce changes to Fe-O-Fe bond angles that affect the superexchange interaction.[64–67] This oxygen-ligand-mediated superexchange enables spins that would have otherwise been misaligned in a high-spin state on the neighboring iron atoms to be transferred to a low-spin $Fe^{2+}$ orbital configuration.[68] However, the bond angle is reduced by ~1.2° according to the DFT calculations, decreasing the superexchange and the effective charge hopping integral.[69–71] The Hubbard interaction is large for the MMCT between Fe centers and electron-phonon coupling can therefore dominate, resulting in polaron formation.

Iron oxides specifically have been reported to experience optical modification of the exchange interaction on sub-picosecond timescales, supporting this conclusion.[68] Particularly in the case of Mott insulators, whose VBM and CBM are dominated by metal orbitals, exchange interactions are calculated to be reversibly modulated by electric fields on ultrafast timescales.[42] The octahedral site distortion in rare-earth orthoferrites is responsible for orbital and spin ordering, the modulation of that distortion through polaron formation enables the trapping of a low-spin $Fe^{2+}$ state induced by Fe-O-Fe superexchange following 800 nm MMCT.[71–73]

# CONCLUSION

Excitation-wavelength-dependent small polaron formation is measured in the intermediate insulator $GdFeO_3$ using transient XUV reflection spectroscopy. Photoexcitation with an 800 nm pump results in both electron and hole polaron formation at neighboring iron sites and strong spectral splitting at the Fe $M_{2,3}$ edge, a spin crossover occurring on the iron atoms. A higher-energy 400 nm pump results in suppression of polarons. This suggests a metal-metal polaron formation mechanism in which nearest neighbor iron sites form a hole and electron polaron. The ability to modulate polaron formation by varying excitation wavelength has been studied previously in hematite.[10] However, this mostly results in changes to the polaron's mobility, lifetime, and recombination properties as carriers populate different energy levels of the conduction bands. In the case of $GdFeO_3$, we have suppressed photoexcited polaron formation with weak on-site repulsions following a higher-energy 400 nm photoexcitation, while inducing polaron formation when electron-phonon coupling is dominated by MMCT following 800 nm photoexcitation. The excitation-wavelength-dependent modulation of polaron formation is of interest for applications that employ polarons for charge separation and for applications that would benefit from carrier mobility offered by the suppression of polaron formation. Tuning spin properties and the exchange integral of these materials present an additional variable for controlling polaron formation in this class of highly polar materials. From a materials engineering perspective, polaronic tunability merits further study in intermediate and Mott-Hubbard insulating transition metal oxide materials.

# ASSOCIATED CONTENT

**Supporting Information**. $GdFeO_3$ characterization, detailed descriptions of the transient XUV spectrometer and ab initio modeling. Long timescale XUV spectra and fitting of polaron shifts. This material is available free of charge via the Internet at http://pubs.acs.org.

# AUTHOR INFORMATION

## Corresponding Author

**Scott K. Cushing** – Division of Chemistry and Chemical Engineering, California Institute of Technology, Pasadena, California 91125, United States

## Authors

**Jocelyn L. Mendes** – Division of Chemistry and Chemical Engineering, California Institute of Technology, Pasadena, California 91125, United States
**Hyun Jun Shin** – Department of Physics, Yonsei University, Seoul, 03722, Republic of Korea
**Jae Yeon Seo** – Department of Physics, Yonsei University, Seoul, 03722, Republic of Korea
**Nara Lee** – Department of Physics, Yonsei University, Seoul, 03722, Republic of Korea
**Young Jai Choi** – Department of Physics, Yonsei University, Seoul, 03722, Republic of Korea
**Joel B. Varley** – Lawrence Livermore National Laboratory, Livermore, California 94550, United States

# NOTES

The authors declare no competing financial interests.

# ACKNOWLEDGEMENTS

The authors thank Professor Hanzhe Liu, Dr. Jonathan Michelsen, and Levi Palmer for guidance and MATLAB scripts for performing the OCEAN calculations. This material is based on work performed by the Liquid Sunlight Alliance, which is supported by the U.S. Department of Energy, Office of Science, Office of Basic Energy Sciences, Fuels from Sunlight Hub under Award Number DE-SC0021266. This research used resources of the National Energy Research Scientific Computing Center a DOE Office of Science User Facility supported by the Office of Science of the U.S. Department of Energy under Contract No. DE-AC02-05CH11231 using NERSC award BES-ERCAP0024109. The computations presented here were, in part, conducted in the Resnick High Performance Computing Center, a facility supported by Resnick Sustainability Institute at the California Institute of Technology. The ground state optical absorption of $GdFeO_3$ was collected at the Molecular Materials Research Center in the Beckman Institute of the California Institute of Technology. Fluorescence lifetime measurements were performed in the Caltech Biological Imaging Center, with the support of the Caltech Beckman Institute and the Arnold and Mabel Beckman Foundation. J.L.M. acknowledges support by the National Science Foundation Graduate Research Fellowship Program under grant no. 1745301. GdFeO$_3$ synthesis and characterization carried out at Yonsei University was supported by the National Research Foundation of Korea (NRF) through grants NRF-2021R1A2C1006375, and NRF-2022R1A2C1006740. The work of J.B.V. was performed under the auspices of the US DOE by Lawrence Livermore National Laboratory under contract DE-AC52-07NA27344 and supported by the HydroGEN Advanced Water Splitting Materials Consortium, established as part of the Energy Materials Network under the U.S. Department of Energy (DOE), the Office of Energy Efficiency and Renewable Energy (EERE), the Hydrogen and Fuel Cell Technologies Office (HFTO).

**TOC**

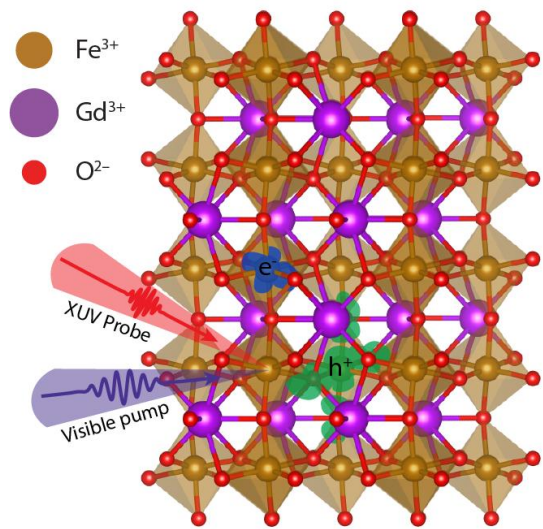

**Supporting Information for:**

# Dynamic Competition between Hubbard and Superexchange Interactions Selectively Localizes Electrons and Holes through Polarons

Jocelyn L. Mendes[1], Hyun Jun Shin[2], Jae Yeon Seo[2], Nara Lee[2], Young Jai Choi[2], Joel B. Varley[3], and Scott K. Cushing[1*]

[1] Division of Chemistry and Chemical Engineering, California Institute of Technology, Pasadena, California 91125, United States
[2] Department of Physics, Yonsei University, Seoul, 03722, Republic of Korea
[3] Lawrence Livermore National Laboratory, Livermore, California 94550, United States
*Correspondence and requests for materials should be addressed to S.K.C. (email: scushing@caltech.edu)

# Supporting Information Contents

## S1. GdFeO₃ Characterization

### S1.1. Ground State Optical Characterization

The band gap of single-crystal GdFeO₃ was determined using UV-Visible reflection spectroscopy (Cary 5000, Agilent Technologies). Scans were collected in a range of 250-1200 nm in reflectivity mode. For a Tauc plot analysis, the reflectance spectra were transformed using the Kubelka–Munk function using **Equation S1**:

$$F(R_\infty) = \frac{(1-R_\infty)^2}{2R_\infty} \quad \textbf{(S1)}$$

where $R_\infty = R_{sample}/R_{standard}$.[1] The direct band gap of GdFeO₃ was measured as ~1.43 eV and the indirect band gap was measured to be ~1.39 eV.

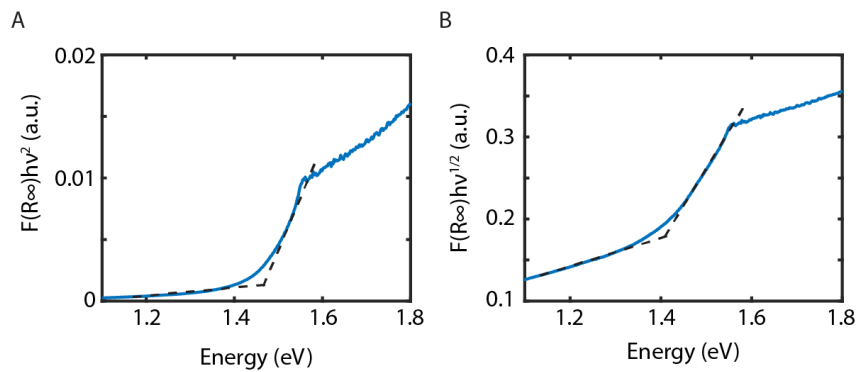

**Figure S1.** Tauc plot from the measured optical ground state reflectivity of GdFeO₃ with $(\alpha h\nu)^{1/r}$ plotted as r = ½ for a direct allowed band gap transition (A) and r = 2 for an indirect allowed band gap transition (B). The direct band gap is found to be ~1.43 eV and the indirect band gap is found to be ~1.39 eV.

### S1.2. Structural Characterization

The crystallographic structure and absence of a second phase were checked by the Rietveld refinement using the FullProf program for the power X-ray diffraction data. The data were obtained with a Rigaku D/Max 2500 powder X-ray diffractometer using Cu-K$_\alpha$ radiation. The result suggests that the GdFeO₃ forms an orthorhombic perovskite with the *Pbnm* space group. The lattice constants are found to be $a$ = 5.3464 Å, $b$ = 5.5996 Å, and $c$ = 7.6623 Å with the reliability factors; $\chi^2$ = 1.64, $R_p$ = 2.29 %, $R_{wp}$ = 3.03 %, and $R_{exp}$ = 2.37 %. Further details of crystallographic data are summarized in **Table S1**.

**Table S1. Crystallographic information of GdFeO₃.** Unit cell parameters, reliability factors and position parameters for GeFeO₃.

| Structure | Orthorhombic |
|---|---|
| Space group | *Pbnm* |
| Lattice parameters (Å) | $a = 5.3464(1)$ |
| | $b = 5.5996(1)$ |
| | $c = 7.6623(2)$ |
| | $c/a = 1.433$ |
| R-factors (%) | $R_p = 2.29$ |
| | $R_{wp} = 3.03$ |
| | $R_{exp} = 2.37$ |
| $\chi^2$ | 1.64 |
| Gd (x, y, z) | (1.0159, 0.0605, 0.25) |
| Fe (x, y, z) | (0.5, 0, 0) |
| O₁ (x, y, z) | (0.4219, 0.9627, 0.25) |
| O₂ (x, y, z) | (0.7107, 0.2988, 0.0455) |

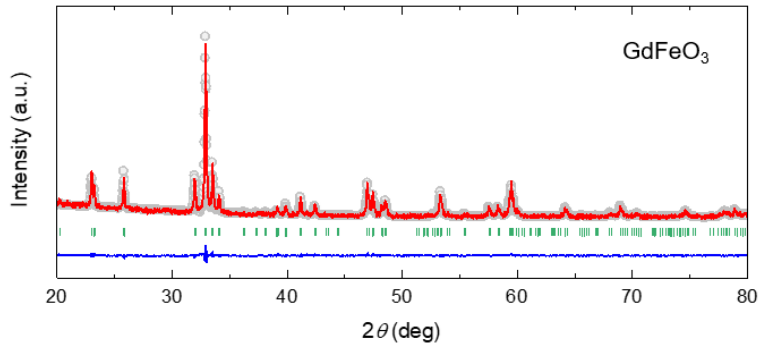

**Figure S2.** X-ray diffraction pattern of single-crystalline GdFeO₃. The observed (gray circles) and calculated (red solid line) powder X-ray diffraction patterns for a GdFeO₃ single crystal at 293 K. The blue curve denotes the difference in intensity between the observed and calculated patterns, while the short green ticks indicate the Bragg reflection positions.

S1.3. Fluorescence Lifetime

The fluorescence lifetime of GdFeO₃ was measured using a Stellaris Falcon Fluorescence Lifetime Imaging Microscope (FLIM, Leica Microsystems). Measurements were performed with the 10x dry lens objective. A CW 405 nm, 0.5 mW laser excites the GdFeO₃ single-crystal, and the resulting fluorescence was measured. The Stellaris Falcon employs TauSense software to gate the PMT detectors with 0.1 ns steps when using CW excitation. Similarly, a 790 nm, 80 MHz beam with 0.5 mW of average power was used to measure fluorescence lifetime. Fitting of lifetimes was conducted using the Leica LAS X software that deconvolves the instrument response function from resulting spectra. The n-exponential reconvolution employed by the Leica LAS X software is shown in **Equation S2**:

$$y(t) = \{IRF(t + Shift_{IRF}) + Bkgr_{IRF}\} \otimes \left\{\sum_{i=0}^{n-1} A[i]e^{\left(-\frac{t}{\tau[i]}\right)} + Bkgr\right\} \quad (S2)$$

Where *n* is the number of exponential components, *A* is the amplitude of exponential pre-factors, $\tau$ is the lifetime, *IRF* is the instrument response function, *Bkgr* is the tail offset correction for the background, *Shift$_{IRF}$* is the correction for the IRF displacement, *Bkgr$_{IRF}$* is a correction for the IRF background, *i* are the intensities associated with each exponential component.

The raw spectra, IRF, and resulting n-exponential reconvolutions for each wavelength can be found in **Figure S3**. The resulting radiative lifetimes calculated using the Leica LAS X software were <100 ps for 790 nm versus 1.2±0.4 ns for 405 nm. We would like to note that the Leica LAS X software calculates a lifetime below 100 ps following 790 nm excitation for the given raw spectra and instrument response function, however, given the limited number of time points in the fluorescence lifetime spectra following 790 nm excitation we refer to it qualitatively as <100 ps.

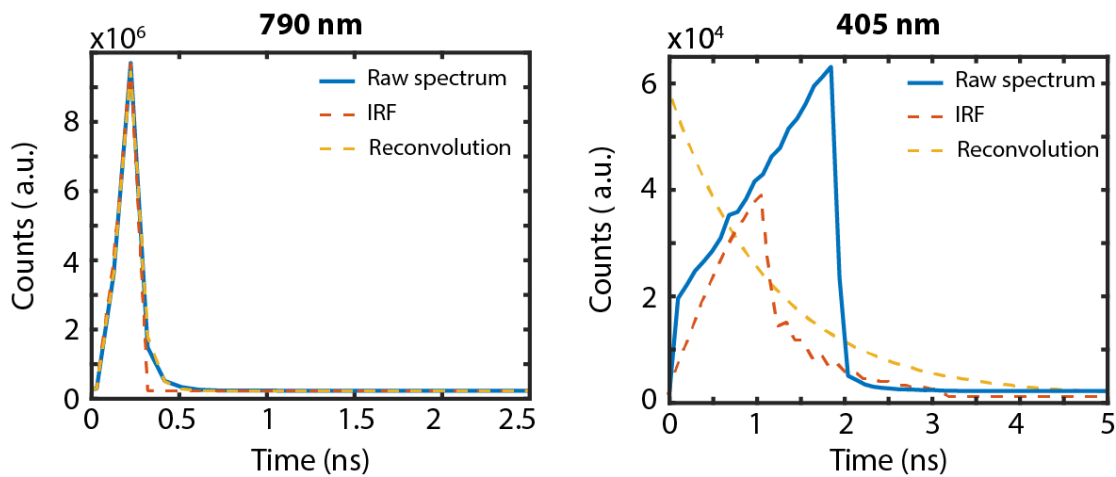

**Figure S3.** Raw fluorescence spectra of GdFeO$_3$ following 790 nm (left) and 405 nm (right) excitation (blue). The red trace denotes the IRF for each wavelength and the yellow trace represents the reconvolution calculated by the Leica LAS X software for the 790 nm and 405 nm.

S2. Experimental Setup: Transient Extreme Ultraviolet Reflectivity Spectrometer

The transient extreme ultraviolet (XUV) reflectivity spectrometer is described previously.[2,3] Briefly, a Legend Elite Duo laser system (Coherent Inc.) with 35 fs, 13 mJ, 1 kHz pulses centered at 800 nm is split by a 75:25 beam splitter with ~3 mJ being used to generate a few-cycle white light beam (<6 fs, 550-900 nm) for high harmonic generation (HHG) and ~10 mJ used for the pump path. The pump path uses both a p-polarized 800 nm beam and the 400 nm frequency doubled output of a BBO crystal with p-polarization, pumped with the 800 nm output of the Ti:Sapphire laser. The optical excitation fluence used in these experiments was ~11 mJ/cm$^2$ for both 400 nm and 800 nm pumped samples. Transient reflection is measured at a 10° grazing incidence (80° from normal incidence) geometry and measures the varying delay times between the pump and probe pulses by an optomechanical delay stage. The photoexcited dynamics were probed with an XUV pulse produced with an s-polarized few-cycle white light pulse by HHG in argon. The residual white light beam is removed with a 200 nm thick Al filter (Luxel). The generated XUV continuum is used to probe the Fe $M_{2,3}$ absorption edges at ~54 eV. An edge-pixel referencing scheme was used to denoise the spectra due to intensity fluctuations and used signal-free spectral regions.[4]

### S2.1. Ground State Reflectivity and Long Timescale Dynamics

The ground state sample reflectivity at the Fe $M_{2,3}$ edge is shown in **Figure S4**. The experimental spectrum (**Figure S4**, orange line) was smoothed using a moving average filter with a span of 2.0% of the pixels in the CCD image. The static XUV reflectivity of $GdFeO_3$ is obtained by normalizing the static XUV reflectivity spectrum of the sample by the static reflectivity of a Si wafer, which should not absorb the XUV beam below the Si K edge at ~100 nm. The spectrometer was calibrated using the photoionization continuum of neon gas.[5]

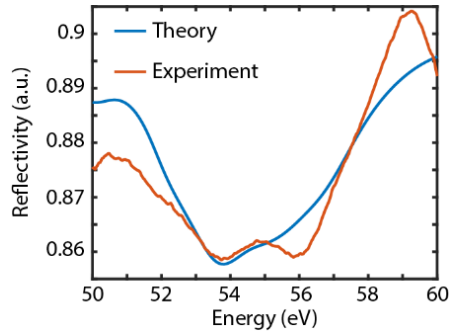

**Figure S4.** Experimental ground state XUV reflectivity spectrum (orange) and DFT+BSE theory calculated ground state reflectivity (blue).

The transient spectra presented in the main text (**Figures 2A and 4**) capture the relevant ultrafast features of charge transfer, polaron formation, and the superexchange induced spin crossover on $Fe^{2+}$ centers. Additionally, we present here representative spectra of long timescale dynamics measured at the Fe $M_{2,3}$ X-ray edge following 800 and 400 nm photoexcitation to understand thermalization in the spectra. **Figure S5A** and **S5B** presents logarithmic spectra collected out to 1 ns. We compare the long timescale 800 and 400 nm pumped spectra to our DFT+BSE treatment of thermal expansion of the $GdFeO_3$ crystal lattice (**Fig. S5C** and **S5D**) and to our modeled charge transfer and polaron states. The polaron state dominates the 800 nm pumped $GdFeO_3$ spectra out to 1 ns (**Fig. S5C**), which suggests that thermal expansion of the lattice plays a minimal role in the observed spectral features. Similarly, the DFT+BSE modeled 400 nm charge transfer state (**Fig. S5D**) continues to dominate the 400 nm pumped transient XUV spectrum, out to 1 ns. This suggests that the free carriers in the charge transfer state dominate the 400 nm pumped spectrum out to our temporal measurement limit.

**Figure S5E** and **F** contain exponential fits of the intensity change with respect to time. The exponential model for both plots followed the equation: $f(x) = a*exp(-x./b)+c$. The resulting non-radiative recombination lifetime for the 800 nm exponential fit was 150±30 ps and for 400 nm excitation it was >1 ns. The radiative lifetime of the 400 nm pumped $GdFeO_3$ discussed in **S1.3** exceeds the 1 ns temporal limit of our XUV spectrometer, resulting in an exponential fit that poorly describes the recombination lifetime of carriers in the 400 nm XUV spectrum.

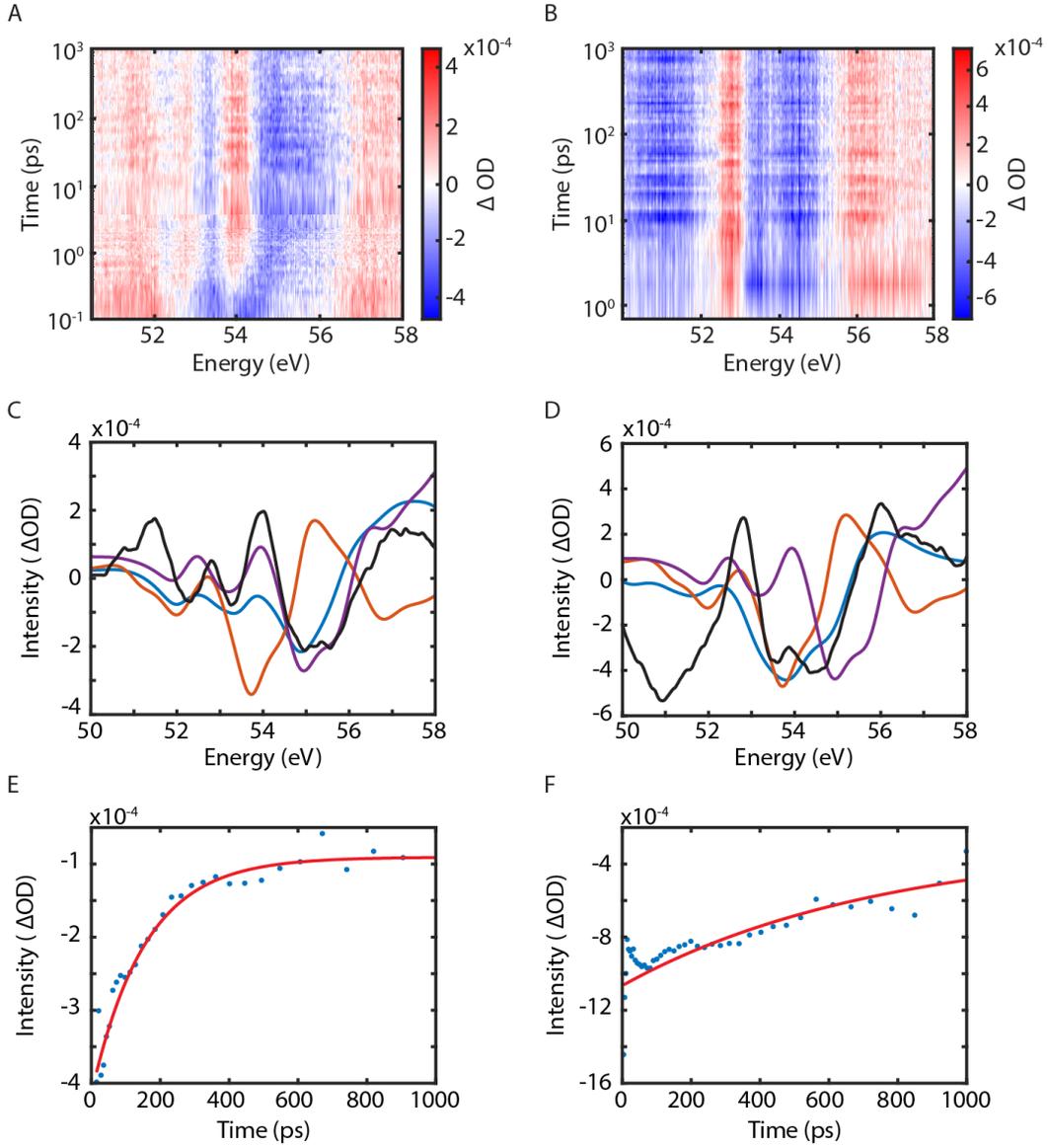

**Figure S5.** Long timescale measurements of GdFeO$_3$ following 800 nm (A) and 400 nm (B) photoexcitation are plotted on a logarithmic time axis. Experimental lineouts 1 ns after 800 nm (C, black) and 400 nm (D, black) photoexcitation are compared to DFT+BSE modeled 800 and 400 nm charge transfer states (blue), the convolved electron and hole polaron state (purple), and a modeled 350 K thermal expansion of the crystal lattice (red). For the 800 nm spectrum (C) the polaron state best agrees with the 1 ns experimental trace, while the 400 nm spectrum (D) agrees well with the 400 nm charge transfer state. Exponential fits of the change in intensity over time for the 800 nm pumped spectrum (E) and the 400 nm pumped spectrum (F).

S2.2. Experimental Photoexcited Carrier Density Approximation

The photoexcited carrier density was calculated assuming that for each photon absorbed by the GdFeO$_3$ single crystal, one electron-hole pair is excited.[6–8] The number of photoexcited carriers ($N_e$) generated by both excitation wavelengths can be approximated by **Equation S3**:

$$N_e = \frac{F}{\hbar v} * \frac{1-R}{D}(1 - \exp(-\alpha D))(1 + R \exp(-\alpha D)) \qquad (S3)$$

where $F$ is the pump fluence (~11 mJ/cm$^2$), $\hbar v$ is the excitation energy, $R$ is the reflectivity coefficient,[9] $D$ is the probe depth (~2 nm), and $\alpha$ is the absorption coefficient.[10] In the case of 800 nm photoexcitation, this results in a photoexcited carrier density of ~7.3x10$^{20}$ cm$^{-3}$, and for 400 nm excitation a carrier density of ~3.7x10$^{20}$ cm$^{-3}$. The DFT calculated unit cell of GdFeO$_3$ has a volume of ~2.4x10$^{-22}$ cm$^3$ and given that each unit cell contains 4 iron atoms, there are approximately ~1.6x10$^{22}$ Fe atoms/cm$^3$. This would result in ~4.5% and ~2.3% of iron atoms excited following 800 nm and 400 nm photoexcitation, respectively.

S3. Spectral Analysis

The spectral shift associated with polaron formation in the 800 nm pumped XUV spectra and the decrease in intensity associated with thermalization of hot carriers at the Fe M$_{2,3}$ edge in the 400 nm pumped spectra were fit with single exponential functions following the equation: a*exp(-x./b)+c. The plots of each can be found in **Figures S6** and **2C** in the main text, respectively. The polaron shift resulted in an exponential fit with a time constant of 250±40 fs and a formation rate of ~4 ps. The spectral shift was measured to have a magnitude of 420±150 meV.

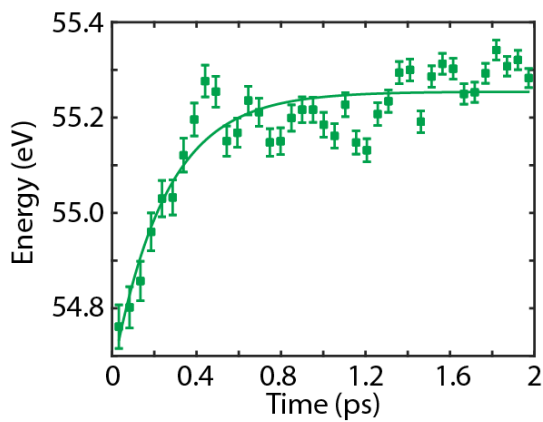

**Figure S6.** A fit of the spectral shift at the Fe M$_{2,3}$ edge following 800 nm photoexcitation. The fitting was performed by finding the minima of the averaged spectral intensity between 54.0 – 55.0 eV.

Additionally, to confirm independent spectral features and kinetics of the resulting phenomena in both 800 and 400 nm pumped XUV spectra, we performed singular value decomposition (SVD). The SVD analysis factorizes the two-dimensional matrix of the transient spectra by rotating (U), scaling (S), and then again rotating (V) the matrix. The scaling matrix contains the singular values along its diagonal, which we refer to as the number of components. The number of singular values or components in the scaling matrix is selected to minimize the total number of singular values.

In the case of 800 nm photoexcitation, two singular values well reproduce the kinetics and spectral features of the transient spectra (**Fig. S7A**). Adding additional singular values to the SVD of the 800 nm transient spectra results in components that look identical to earlier components, but with lower intensity contributions, suggesting that two is sufficient and any additional components are negligible. The SVD components for the 800 nm spectra plotted in **Figure S7A** agree well with both the polaron state (component 1) and the MMCT state (component 2). To visualize the kinetics of the components, **Figure 5C** in the main text, plots the components associated with the MMCT state and the polaron state as a function of time using the experimental 800 nm transient spectrum.

For the SVD of the 400 nm pumped XUV spectra, one component is sufficient to successfully reproduce the transient spectrum. Suggesting that one process dominates the spectra and kinetics of the plot. The component in **Figure S7B** agrees well with the 400 nm LMCT state, suggesting that aside from carrier thermalization, this state dominates the 400 nm transient kinetics.

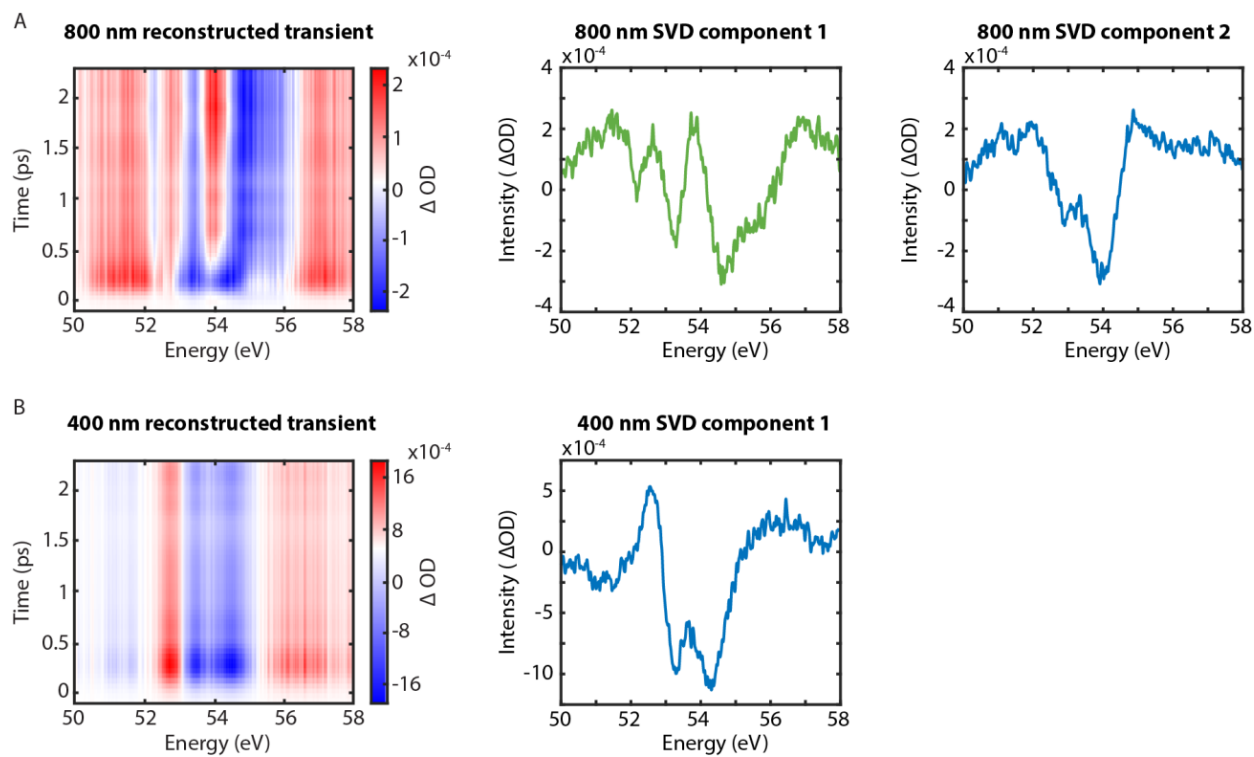

**Figure S7.** Singular value decomposition analysis of 800 nm (A) and 400 nm (B) pumped XUV spectra. Decomposing the 800 nm pumped XUV spectra into two components results in a reconstruction of the transient spectra (A, left) that agrees well with the experimental spectra. The first component of the 800 nm SVD analysis (A, middle) agrees well with the polaron state, and the second component (A, right) agrees well with the 800 nm MMCT state. The 400 nm pump spectra can be reconstructed (B, left) in good agreement with the experimental spectra using only one component (B, right), which agrees well with the 400 nm LMCT state.

S4. Theoretical Methods: Ground State and Excited State Core-Level Spectra
   S4.1. Density Functional Theory Calculations

Geometry optimization and density functional theory (DFT) were conducted using the Quantum ESPRESSO package with norm-conserving general gradient approximation (GGA), Perdew–Burke–Ernzerhof (PBE) pseudopotentials.[11–13] Geometry optimizations employed an input $GdFeO_3$ unit cell containing 20 atoms and based upon the experimental crystallographic information listed in **Table S1**. A 250 Rydberg kinetic energy cutoff and an $11 \times 10 \times 7$ Monkhorst–Pack k-mesh sampling of the first Brillouin zone was used to sample 200 bands. The starting magnetization was set to a sum of zero for $Fe^{3+}$ ions to represent the antiferromagnetic orientation of neighboring spins at 293 K.[9] A Hubbard $U$ parameter of 4 eV and 5 eV was applied to Fe and Gd atoms respectively for DFT and DFT+BSE calculations. The structures employed in the DFT and DFT+BSE calculations described below were fully relaxed and calculations were performed to a convergence of $10^{-8}$ eV/atom with the forces on ions under $10^{-3}$ eV/Å.

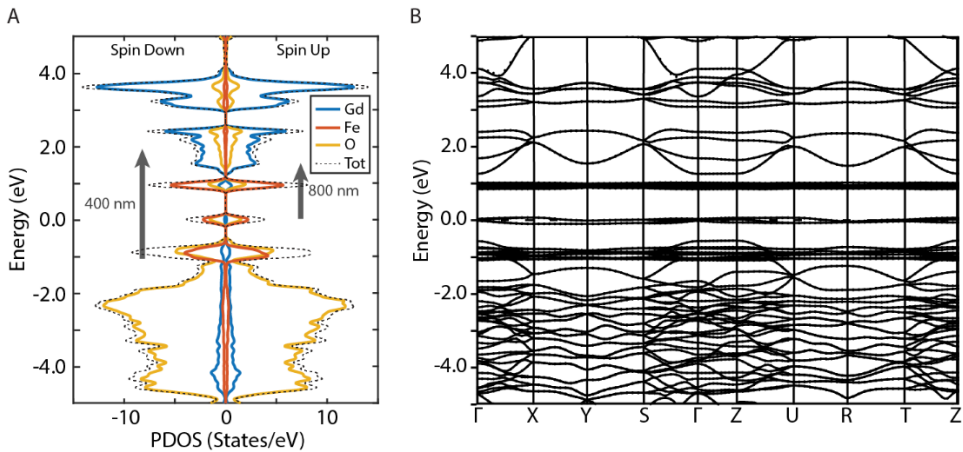

**Figure S8.** DFT+U calculated projected density of states (A), and band structure (B) used for DFT+BSE calculations of core-level spectra.

### S4.2. Defect Supercell Calculations of Polarons

Hole and electron polarons were modeled within a supercell approach,[14] using a $2 \times 2 \times 1$ repetition of the optimized unit cells (160-atom supercells) calculated for an antiferromagnetic spin ordering ground state. The defect supercell calculations with pSIC and hybrid functionals were performed using the VASP code version 5.4,[15] using the HSE06 screened hybrid functional,[16,17] the implemented pSIC approach,[18,19] and the projector augmented wave (PAW) approach with the VASP4 PAW_PBE potentials (Fe_pv : 14 valence electrons, Gd: 18 valence electrons, and O: 6 valence electrons).[20,21] All supercell simulations adopted a plane-wave energy cutoff of 400 eV, and evaluated energies with a single special $k$-point at 0.25, 0.25, 0.25. Simulations with DFT+U with pSIC adopted a Hubbard $U$ value only for the Fe d states with a value of 5.3 as based on default parameters used for the Materials Project database.[22,23] All initial structures were optimized from unit cells with an antiferromagnetic ground state with magnetic moments of 4.8 and 7.2 for the Fe and Gd, respectively, with final moments of approximately 4 and 7 obtained for the bulk for the studied levels of theory. The lattice constants for the supercells with self-consistency determined for each level of theory, with the impact of the fraction of exact change on the electronic band included in **Figure S9**.

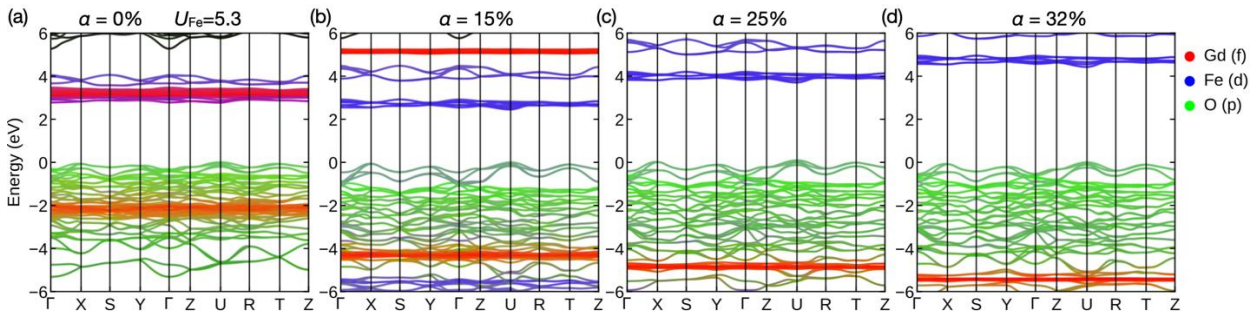

**Figure S9.** Summary of variations in the electronic structure for GdFeO$_3$, shown at different levels of theory from PBE+U (a) to HSE06 with different levels of exact exchange such as 15% (b), 25% (c) and 32% (d). The primary impact of exact exchange is widening of the band gap between the upper valence band (Fe and O derived) and the localized Fe-d derived band, with the separation between the latter band and a higher-lying conduction band (Fe and Gd derived) remaining largely ~ 1 eV higher, independent of the exact exchange.

### S4.3. Fe M$_{2,3}$ Edge Ground State Calculation

The fully relaxed parameters used in the DFT+U calculations for GdFeO$_3$ discussed in **S4.1** were applied to the DFT+BSE calculations of the GdFeO$_3$ ground state reflectivity at the Fe M$_{2,3}$ X-ray edge. The BSE calculation employed a screening mesh of $2 \times 2 \times 2$, a dielectric constant of 4.28 at 293 K,[9] a 4.0 Bohr screening radius, and a 0.7 scaling factor for the Slater G parameter. A spectral broadening of 0.25 eV was applied for all calculated spectra. The calculated ground state reflectivity of GdFeO$_3$ is plotted in **Figure S4** and is compared to the experimentally measured XUV ground state at the Fe M$_{2,3}$ X-ray edge. As shown in **Figure S4**, the DFT+BSE calculated ground state agrees well with the measured ground state reflectivity of the GdFeO$_3$ sample.

### S4.4. Fe M$_{2,3}$ edge Excited and Thermal State Calculations

Excitation-wavelength-dependent changes to the transient XUV spectra of GdFeO$_3$ were calculated using a modification to the OCEAN package described previously.[3,24–26] Effectively, the modification to the OCEAN code enables us to selectively allow or forbid core-to-valence X-ray transitions to valence and conduction bands at different points in momentum space, to simulate state-filling of electrons and holes following photoexcitation. We modeled our charge transfer state by populating the conduction band with electrons either at 1.55 eV (800 nm) or 3.1 eV (400 nm) above the valence band maximum (VBM) and similarly populating the valence band with holes to model the movement of carriers following photoexcitation. This population of electrons and holes effectively forbids or allows XUV transitions to specific points in k-space but does not account for carrier density, therefore, the modeled peak position and relative intensity is used to compare experiment to theory. We model our transient XUV spectra by subtracting the calculated ground state spectrum from the calculated excited state spectra to calculate the differential absorption. **Figures 2 and 5** in the main text demonstrate agreement between DFT+BSE modeled 400 nm and 800 nm charge transfer states, respectively.

A thermal isotropic expansion of the GdFeO$_3$ unit cell was modeled to determine if thermal effects were present in the experimental spectra and is described previously.[27] We isotropically expanded the GdFeO$_3$ unit cell to model temperatures from 293 K up to 500 K.[28] The spectrum of a modeled 350 K expansion can be seen in **Figure S5C** and **D**. We find no significant contributions of thermal isotropic lattice expansion to 800 nm or 400 nm excited transient XUV spectra.

### S4.5. Fe M$_{2,3}$ Edge DFT+BSE Polaron State Calculations

Our semi-empirical method for modeling transient effects of photoexcited polaron formation on XUV core-level spectra is achieved using our DFT+BSE framework and has been described previously.[27]

Electron polarons are modeled using the bond distortion method,[29] locally applying an expansion to Fe–O bonds at one iron octahedra in the GdFeO$_3$ unit cell. Similarly, hole polarons on iron atoms are modeled by applying a local contraction to the Fe–O bonds in an FeO$_6$ octahedra. **Figure S10** demonstrates the change in intensity between a modeled electron polaron and hole polaron centered on an iron atom. Both polarons were modeled with a 1-8% isotropic distortion of the local FeO$_6$ octahedra, with a 5% distortion agreeing best with experimental spectra. The electron polaron has a higher spectral intensity than the more dispersive hole polaron. We also simultaneously model electron and hole polarons with the bond distortion method at neighboring iron sites by employing an expansion and contraction to consider spectrally convolved electron and hole polarons at the Fe M$_{2,3}$ edge (**Fig. S10**). The electron and hole polarons on iron centers are spectrally convolved at the Fe M$_{2,3}$ edge, and occur at the same energy alignment, which is shifted to higher energy relative to both the LMCT and MMCT states. These anisotropic distortions to the GdFeO$_3$ unit cell are then applied to our DFT+BSE method and the resulting XUV spectrum is treated similarly to our excited state calculation detailed in **S4.4** to generate a differential absorption spectrum (**Fig. S10**).

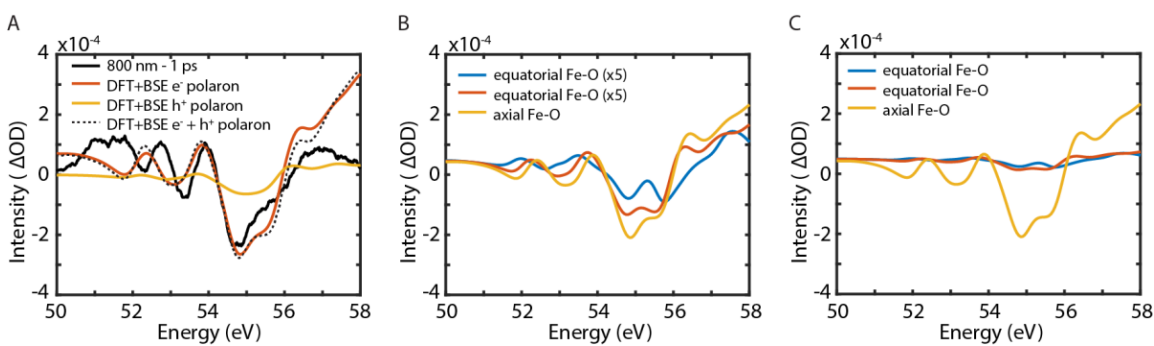

**Figure S10.** (A) Experimental lineout (black solid) 1 ps after 800 nm photoexcitation compared to DFT+BSE modeled electron (red) and hole (yellow) polarons and compared to a convolved DFT+BSE electron and hole polaron at neighboring iron atoms (black dashed). The DFT+BSE modeled hole polaron is much weaker than the electron polaron due to its delocalized nature, and when convolved together the electron polaron dominates the predicted dynamics. (B) A DFT+BSE modeled electron polaron for an anisotropic expansion of either the equatorial (blue and red) or axial (yellow) Fe–O bonds. The equatorial lineouts are each scaled by 5 times to be compared to the axial distortion. The equatorial ligand expansion is significantly weaker than the axial ligand expansion as plotted in (C) and demonstrates that axial distortions dominate the spectral signature of polaron formation.

Additionally, we model an anisotropic polaronic distortion to stimulate the Jahn–Teller effect. To achieve this distortion, we selectively distort equatorial or axial ligands in the FeO$_6$ octahedra. **Figure S10B and C** shows that an anisotropic polaronic expansion results in different spectral contributions to the signature of the polaron. We find that axial distortions have a greater effect on the spectral intensity of the polaronic feature in the transient XUV spectrum, which agrees with our findings that the polaronic distortion induces a Jahn–Teller type distortion in the lattice.

### S4.6. Fe M$_{2,3}$ Edge pSIC+DFT+BSE Polaron State Calculations

In addition to our semi-empirical modeling of polaronic distortions, we apply the polaronic distortions obtained from the ab initio defect supercell calculations discussed in **S4.2** to our DFT+BSE approach, which we refer to as pSIC+DFT+BSE. The distortion acts as the input for OCEAN and enables us to simulate the excited state X-ray edges that result from the ab initio electron and hole polaron distortions calculated using the defect supercell method.

### S5. Ligand-Field Multiplet Theory Calculations

Ligand-field multiplet theory calculations (CTM4XAS) were conducted to model the XUV absorption of oxidation and spin states of different iron species.[30] The different oxidation states were modeled by designating the final state of the initial $Fe^{3+}$ state to be either $Fe^{2+}$ or $Fe^{4+}$. Spin configurations of the photoexcited $Fe^{2+}$ and $Fe^{4+}$ states were modeled by varying the cubic crystal field splitting (10Dq) and the exchange energy ($J$) for either configuration. The $Fe^{3+}$ high-spin ground state was subtracted from the $Fe^{2+}$ and $Fe^{4+}$ excited states to generate predicted difference spectra following photoexcitation. **Figure S11** compares these difference spectra with the experimental 400 nm LMCT, 800 nm MMCT, and 800 nm polaron states. The ground state bleach of the $Fe^{3+}$ high-spin feature was shifted to align with experimental bleach at the Fe M$_{2,3}$ edge. The 400 nm LMCT state agrees with a high-spin $Fe^{2+}$ state (**Fig. S11A**) and the 800 nm polaron state (**Fig. S11C**) agrees well with a $Fe^{2+}$ low-spin and $Fe^{4+}$ high-spin configuration.

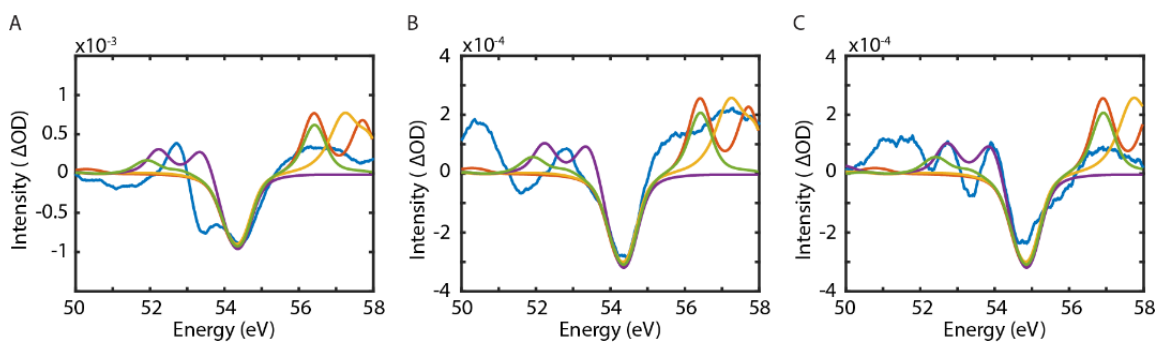

**Figure S11.** (A) 400 nm LMCT state, (B) 800 nm MMCT state, (C) 800 nm polaron state. Experimental (blue), $Fe^{4+}$ high-spin state (yellow), $Fe^{4+}$ low-spin state (red), $Fe^{2+}$ high-spin state (green), $Fe^{2+}$ low-spin state (purple).